\documentstyle[12pt]{article}
\begin{document}
\hspace {10.cm} DFUPG 100-95\\
\vspace {1.5cm}
\begin{center}
{\large\bf Confinement-Deconfinement Transition in 3-Dimensional
QED}\\
\vspace {1 cm}
{\large G. Grignani\footnote{Permanent address: Dipartimento di Fisica
and Sezione I.N.F.N., Universit\'a di Perugia, Via A. Pascoli I-06100
Perugia, Italy}, G. Semenoff and P. Sodano$^1$}\\
\vspace {0.4cm}
Department of Physics,
University of British Columbia\\Vancouver, British Columbia, Canada V6T 1Z1
\\
\vspace {0.4cm}
\end{center}
\vspace {1 cm}
\centerline{\bf Abstract}
\vspace {0.4 cm}
We argue that, at finite temperature, parity invariant non-compact
electrodynamics with massive electrons in 2+1 dimensions can exist in
both confined and deconfined phases.  We show that an order parameter
for the confinement-deconfinement phase transition is the Polyakov
loop operator whose average measures the free energy of a test charge
that is not an integral multiple of the electron charge.  The
effective field theory for the Polyakov loop operator is a
2-dimensional Euclidean scalar field theory with a global discrete
symmetry $Z$, the additive group of the integers.  We argue that the
realization of this symmetry governs confinement and that the
confinement-deconfinement phase transition is of
Berezinskii-Kosterlitz-Thouless type.  We compute the effective action
to one-loop order and argue that when the electron mass $m$ is much
greater than the temperature $T$ and dimensional coupling $e^2$, the
effective field theory is the Sine-Gordon model.  In this limit, we
estimate the critical temperature, $T_{\rm crit.}=e^2/8\pi(1-e^2/12\pi
m+\ldots)$.
\newpage

Gauge field theories in 2+1-dimensions have many interesting field
theoretical features such as gauge invariant local topological
mass~\cite{djt2,tpm}, fractional spin~\cite{3.} and statistics
\cite{4.}. They have been of interest as model systems with somewhat
more severe infrared divergences than their 3+1-dimensional
relatives~\cite{djt2,ap} and where one can study the dynamics of chiral
symmetry breaking~\cite{pisarski,abw,semsod}.  Gauge field theories
arise naturally in the description of lower dimensional statistical
models such as spin systems and the Hubbard model~\cite{fra} and an
understanding of both their ground state and thermodynamic properties
are essential to the applications of mean field theory there.  In this
Letter, we shall show that parity invariant 2+1-dimensional quantum
electrodynamics exhibits one more interesting property - when the
electron has a large enough mass - it has a finite temperature
confinement-deconfinement phase transition.

The question of confinement has been extensively investigated in the
more familiar context of quantized Yang-Mills theory at finite
temperature and it is intimately related to the realization of a global
symmetry involving the center of the gauge group~\cite{po,su}.  This
symmetry transforms the Polyakov loop operator
\begin{equation}
P(\vec x)\equiv {\rm tr}{\cal P}\exp\left(\int_0^{1/T}d\tau A_0(\tau, \vec
x)\right)
\end{equation}
which measures gauge group holonomy in the periodic Matsubara time in
the Euclidean path integral description of finite temperature gauge
theory.\footnote{We use units where Planck's constant,
the speed of light and Boltzmann's constant are one.
For a discussion of the path integral formulation
of finite-temperature gauge theory, see~\cite{gpy}.}
Consider a gauge transformation
\begin{equation}
{A_\mu}'(\tau,\vec x)= g^{-1}(\tau,\vec x)A_\mu(\tau,\vec
x)g(\tau,\vec x)+ig^{-1}(\tau,\vec x)
\nabla_\mu g(\tau,\vec x)
\label{gauge}
\end{equation}
under the group $SU(N)$ which has center $Z_N$,
the additive group of the integers modulo $N$.
In Eq.(\ref{gauge}) $g(\tau,\vec x)$ can be periodic up
to an element of the center of the group, $g({1/T},\vec x)=g(0,\vec
x)e^{2\pi i n/N}$. Under a gauge transformation of
the form (\ref{gauge}),
\begin{equation}
P'(\vec x)=P(\vec x)e^{2\pi in/N}
\end{equation}
Therefore, if the $Z_N$ symmetry is not spontaneously broken, the
correlators of Polyakov loop operators $<P(\vec x_1)\ldots P(x_m)
P^{\dagger}(\vec y_1) \ldots P^{\dagger}(\vec y_n)>$, are zero unless
$m=n$ modulo $N$. The quantity
\begin{equation}
F(\vec x_1,\ldots,\vec x_m,\vec y_1,\ldots,\vec y_n)=
-T\ln\left(<P(\vec x_1)\ldots P(x_m)P^{\dagger}(\vec y_1)
\ldots P^{\dagger}(\vec y_n)>\right)
\end{equation}
is the free energy of the gluodynamic system in the presence of an
external fundamental representation quark sources at positions $\vec
x_1, \ldots \vec x_m$ and anti-quark sources at $\vec y_1,\ldots,\vec
y_n$.  If the $Z_N$ symmetry is unbroken, the expectation value of a
single loop, $<P(\vec x)>$, vanishes and consequently the free energy
$F(\vec x)$ of a single quark source is infinite.  This is a signal of
confinement - introducing a colored source into the confining system
requires infinite energy.  On the other hand, if the $Z_N$ symmetry is
spontaneously broken, $F(\vec x)$ is finite and characterizes the
deconfined phase~\cite{bgkp}.

When dynamical quarks in the fundamental representation of the gauge
group are present, the Polyakov loop operator does not characterize
the confining phase, since, even if quarks are confined, screening can
still take place. The coupling of gluons to quarks which are in the
fundamental representation, is invariant only under strictly periodic
gauge transformations and therefore breaks the $Z_N$ symmetry
explicitly.  This is interpreted as the possibility of fundamental
quarks screening the color of an external source, so that the free
energy of the source is always finite.

In this Letter, we shall argue that the Polyakov loop operator can be
used to study confinement in non-compact quantum electrodynamics even
when dynamical electrons are present.  In this case, the free energy
of a distribution of external charges is
\begin{equation}
e^{- F(\vec x_i)/T}={\int dA_\mu d\psi d\bar\psi e^{-\int_0^{1/T}(
{1\over4}F_{\mu\nu}^2+\bar\psi(\gamma\cdot(\nabla-ieA)+m)\psi)}
e^{i\sum e_i \int_0^{1/T} d\tau A_0(\tau,\vec x_i)} \over
\int dA_\mu d\psi d\bar\psi e^{-\int_0^{1/T}(
{1\over4}F_{\mu\nu}^2+\bar\psi(\gamma\cdot(\nabla-ieA)+m)\psi)} }
\label{(2)}
\end{equation}
with (anti-)periodic boundary conditions $A_\mu({1/T},\vec
x)=A_\mu(0,\vec x)$, $\psi({1/T},\vec x)=-\psi(0,\vec x)$,
$\bar\psi({1/T},\vec x)=-\bar\psi(0,\vec x)$.  The gauge
transformation
\begin{equation}
{A_\mu}'(\tau,\vec x)= A_\mu(\tau,\vec x)+\nabla_\mu\chi(\tau,\vec x)
{}~,~\psi'(\tau,\vec x)= e^{ie\chi(\tau,\vec x)}\psi(\tau,\vec x)
{}~,~\bar\psi'(\tau,\vec x)= \bar\psi(\tau,\vec x) e^{-ie\chi(\tau,\vec
x)}
\label{(3)}
\end{equation}
is a symmetry of the action and measure if it preserves the
(anti-)periodic boundary conditions, $\vec\nabla_\mu\chi(1/T,\vec
x)=\nabla_\mu\chi(0,\vec x)$ and $\chi({1/T},\vec x)=\chi(0,\vec
x)+2\pi n/e$.  The coset of the group of all time-dependent gauge
transformations modulo those which are periodic is the group $Z$, the
additive group of the integers.  The Abelian Polyakov loop operator
transforms under this global symmetry as
\begin{equation}
\exp\left({i\sum_i e_i \int_0^{1/T} d\tau {A_0}'(\tau,\vec x_i)}\right)=
\exp\left({i\sum_i e_i \int_0^{1/T} d\tau A_0(\tau,\vec x_i)}\right)\exp\left(
{{2\pi in\over e}\sum_i e_i}\right)
\label{(4)}
\end{equation}
Thus, if $Z$ is not spontaneously broken, $F(\vec x_i)$ defined by
(\ref{(2)}) is infinite when the total charge of the external
distribution is not an integral multiple of the electron charge,
$\sum_i e_i\neq{\rm integer}\cdot e$.  When the symmetry is broken,
$F(\vec x_i)$ can be finite.  Thus, the nature of the realization of
$Z$ tests the ability of the electrodynamic system to screen charges
which are not integral multiples of the electron charge and therefore
probes the confining nature of the electromagnetic interaction.

$Z$ is the analog of the global $Z_N$ symmetry of gluo-dynamics.
However, in contrast to the case of a compact non-Abelian gauge
theory, where $Z_N$ is explicitly broken by dynamical quarks, and, due
to compactness of the gauge group, charges which are not integer
multiples of the quark charge are not available, the non-compactness
of the gauge group of QED allows the $Z$ symmetry to exist even in the
presence of dynamical electrons. The electrons could be viewed as the
analog of adjoint particles, either gluons or adjoint quarks, in QCD.

At $T=0$, and for the physical value of the electromagnetic coupling
constant, 3+1-dimensional electrodynamics does not exhibit a confining
phase.  It is in the deconfined Coulomb phase at zero temperature and
forms a Debye plasma at finite temperature and density.  There is a
conjecture that, if the electron charge is increased so that
$e^2/4\pi\sim 1$, there is a phase transition to a chiral symmetry
breaking and perhaps confining phase ~\cite{mi}.  This phase, being in
the strong coupling region, is difficult to analyze.  In
1+1-dimensions, the massive Schwinger model is confining and the $Z$
symmetry is not broken, at least when the temperature is much greater
than the electron mass and the confinement scale is set by the
dimensional electron charge $e$.  On the other hand, it is known that
when the electron mass is zero, the $Z$ symmetry is spontaneously
broken ~\cite{hnz}.  The symmetry breaking can be attributed to
nonlocal effects of massless fermions.  It can be argued that the
phase transition between the broken $Z$ and unbroken $Z$ phases occurs
for all temperatures at some value of the electron mass.

One might also expect a confinement-deconfinement transition in the
intermediate case of $2+1$-dimensional electrodynamics.  In that case,
even at the classical level, the Coulomb potential is a marginally
confining logarithm. Its entire spectrum is bound states, but the
bound states can have arbitrarily large size.  The free energy of a
gas of charged particles is
\begin{equation}
F_{\rm cl.}=-\frac{1}{2}\sum_{i,j}e_ie_j\frac{1}{2\pi}\ln\vert\vec x_i
- \vec x_j \vert
\end{equation}
It is straightforward to compute the correlators of Polyakov loop
operators (which are simply exponentials of the appropriate free
energies, $e^{-F_{\rm cl.}/T}$).  They have the scaling form
\begin{equation}
<\prod_i e^{ie_i\int_0^{1/T}d\tau A_0(\tau,\vec x_i)}>={\rm const.}
\prod_{i<j}~\vert \vec x_i-\vec x_j\vert^{e_ie_j/2\pi T}
\end{equation}
with temperature dependent exponent, reminiscent of the spin-wave
correlators in Gaussian spin wave theory in 2 dimensions~\cite{gold}.

It is interesting to ask how this result would be changed by radiative
corrections and by thermal fluctuations.  This can be done by
computing the effective action for the Polyakov loop operator.  In
electrodynamics at finite temperature, it is possible to use a gauge
transformation to set the temporal component of the gauge field,
$A_0$, independent of the Euclidean time.  Then, the effective
2-dimensional field theory for $A_0$ is obtained by integrating the
other degrees of freedom from the path integral.  What remains is an
effective action for a static field $A_0(\vec x)$.  In general, it was
shown in ref.~\cite{gss} that the effective action for the Polyakov
loop operator in D+1-dimensional gauge theory is a D-dimensional sigma
model with group-valued fields.  In the present case of
electrodynamics, the group-valued variables are $e^{ieA_0/T}$ and the
effective field theory for $A_0$ describes the appropriate sigma
model.  The $Z$ symmetry is a periodicity of the effective action
under the field translation $A_0(\vec x)\rightarrow A_0(\vec x)+ 2\pi
T/e$.

The effective action obtained from integrating propagating fields from
the path integral is non-local and non-polynomial in the remaining
fields.  It can only be regarded as a local field theory when the
momenta of interest are much smaller than the masses of the fields
which have been eliminated.  In that case the effective action has a
local expansion in powers of derivatives divided by masses.  The
effective action for $A_0$ possesses such a local expansion. If the electron
mass is sufficiently large that this expansion is accurate, the
effective field theory for $A_0$ can be approximated by a local field
theory.

In 2+1-dimensions, the fermion mass operator constructed from the
minimal 2-component Dirac fermions is a pseudoscalar and therefore
violates parity ~\cite{djt2,djt}.  If included in the action, they
can generate a parity violating topological mass for the photon by
radiative corrections ~\cite{ns,re}.  In this paper, we wish to study
the case where the photon is massless.  For this purpose, we study the
model with Euclidean action
\begin{equation}
S=\int d^3x\left[
\frac{1}{4}F_{\mu\nu}^2+\bar\psi_1(\gamma\cdot(\nabla+ieA)+m)\psi_1
+\bar\psi_2(\gamma\cdot(\nabla+ieA)-m)\psi_2\right]
\label{qed}
\end{equation}
Where the parity transformation is the spacetime parity as well as
$\psi_{1,2}'(x')= \gamma_1\psi_{2,1}(x)$,
$\bar\psi_{1,2}'(x')=-\bar\psi_{2,1}(x)\gamma_1$ with
$x'=(-x_1,x_2)$.  The mass term
$\bar\psi_1\psi_1-\bar\psi_2\psi_2$ is a scalar. If $m=0$ in
(\ref{qed}) there is a `chiral' symmetry under the transformation
${\psi_i}'(x)=u_{ij}\psi_j(x)$ where $u\in~SU(2)$.  The latter
symmetry is broken at $T=0$~\cite{pisarski,abw}, however, since it is
a continuous symmetry, it must be unbroken at any finite temperature
in 2+1-dimensions.  It has been argued that at finite temperature, the
chiral transition is replaced by a Berezinskii-Kosterlitz-Thouless
(BKT) transition ~\cite{ros}.  In this paper, we shall consider the
opposite limit of large mass, and show that there is indeed a
BKT-transition corresponding to a confinement-deconfinement transition
at some value of the temperature $T$.  It is likely that this
transition is in some way related to the chiral transition.

At finite temperature, 2+1-dimensional QED contains three parameters
with the dimension of mass, the electron mass $m$, the gauge coupling
$e^2$, and temperature $T$.  The loop expansion is
super-renormalizable ~\cite{djt2,ap} and is an expansion in the
dimensionless ratios $e^2/m$ and $e^2/T$.  We can compute the
effective action for $A_0(\vec x)\equiv a(\vec x)\sqrt{T}$ in a double
expansion in the number of loops and in powers of derivatives of
$a(\vec x)$.  To order 1-loop and up to quadratic order in derivatives
the effective action is
\begin{equation}
S_{\rm eff}[A_0]=\int d\vec x \left( Z(m,ea/{\sqrt{T}})
\frac{1}{2}\vec\nabla a\cdot\vec\nabla
a -V(m,ea/\sqrt{T})\right)\ \ .
\label{effa}
\end{equation}
Here $V$ is the effective potential for $A_0$ arising from the fermion
determinant and $Z$ is obtained from expansion of the temporal
components of the vacuum polarization function to linear order in
$-\vec\nabla^2$. To order one-loop, the effective potential is
obtained from the fermion determinant in constant background $A_0$,
\begin{equation}
V(m,eA_0/T)=\frac{1}{({\rm Vol.})}
\log \det((-i\partial_0-e A_0)^2-\nabla^2+m^2)
\end{equation}
where the fermions have anti-periodic boundary conditions in the $0$
direction.  The determinant can be computed by considering the ratio
{}~\cite{bal}
\begin{equation}
\Delta(m,eA_0/T)=\det((-i\partial_0-e
A_0)^2-\nabla^2+m^2)/ \det(-\partial^2_0-\nabla^2+m^2)
\end{equation}
One finds
\begin{equation}
\Delta(m,e A_0/T)=\prod_{\vec k}\left[1-\frac{\sin^2(e
A_0/2T)}{\cosh^2(\lambda_k/2T)}\right]\equiv\prod_{\vec k}\Delta_{\vec
k}(m,e A_0/T)\ \ , \label{det}
\end{equation}
where $\lambda_k^2=\vec k^2+m^2$ are the positive eigenvalues of the
operator $-\nabla^2+m^2$. Eq.(\ref{det}) holds in any dimensions. In
$2+1$-dimensions however one can perform the integral on $\vec k$
arising in $\log\Delta(m,eA_0/T)$, after taking the infinite volume
limit.
\begin{eqnarray}
V(m,eA_0/T)=\frac{1}{({\rm Vol.})}\int_{-\infty}^{+\infty}
\frac{d^2\vec{k}}{(2\pi)^2}\log
\Delta_{\vec k}(m,eA_0/T)=~~~~~~~~~~~~~~~\nonumber \\
=-\frac{T^2}{\pi}\left[\frac{m}{T}Li_2(e^{-m/T},eA_0/T+\pi)+
Li_3(e^{-m/T},e A_0/T+\pi)\right]
\label{eff}
\end{eqnarray}
where $Li_2(r,\theta)=-\int_0^r dx \ln(1-2x\cos\theta+x^2)/2x$ and
$Li_3(r,\theta)=\int_0^r dx Li_2(x,\theta)/x$ are the real parts of
the dilogarithm and trilogarithm according to the convention of
Ref.~\cite{lew}.  Eq.(\ref{eff}) shows the periodicity of the
effective potential for $e A_0/T\to e A_0/T+2\pi$. This is the
residual gauge invariance.  In 1 and 3 dimensions the integral in
$\vec{k}$ can only be performed for $m=0$ in which case it gives
simple polynomial expressions. In the limit $m=0$, the effective
potential for $A_0$ has been discussed in ~\cite{sm}.

It is also straightforward to compute the term which contributes the
leading order in derivatives to the effective action,
\begin{equation}
Z(m,e a/\sqrt{T})=\frac{e^2}{12\pi m}\left(\frac{\sinh m/T}{\cosh m/T+
\cos ea/\sqrt{T}}-\frac{m}{T}\frac{e^{-m/T} }{\left(\cosh m/T+
\cos ea/\sqrt{T}\right)^2}\right)
\label{effb}
\end{equation}

The critical behavior of the 2-dimensional model
defined by Eqs.(\ref{effa}), (\ref{eff}) and (\ref{effb}), can be understood by
comparing it with the sine-Gordon model in two-dimensions.  That this
comparison
can be reliably performed can be seen by the study of the harmonic
content of (\ref{eff}).
\begin{equation}
V(m,e a/\sqrt{T})=-\frac{T^2}{\pi}\sum_{n=1}^{\infty}e^{-n m/T}
\left(1+\frac{n m}{T}\right)\cos(n(e a/\sqrt{T}+\pi))\ \ .
\end{equation}
Consider then the large $m$ limit, $T/m$ and $e^2/m$ small with finite
$e^2/T$.  In this limit, the higher harmonics are small perturbations
to the potential
\begin{equation}
V(m,e a/\sqrt{T})=\frac{T m}{\pi}e^{-m/T}\cos(e a/\sqrt{T})\ \ ,
\label{sing}
\end{equation}
which is the sine-Gordon potential.  Amit et al. ~\cite{am} showed
that in the sine-Gordon model any perturbation of the type
$\cos(n\beta\phi)$ to a sine-Gordon potential
$\alpha\cos(\beta\phi)/\beta^2$ are irrelevant for the critical behavior
of the model.  By analogy with the spin wave plus Coulomb gas model,
it was also proven in Refs.~\cite{am} that a critical line for a BKT
{}~\cite{ber,kos} phase transition in the sine-Gordon model with a
logarithmic potential starts at the point $(\alpha,\beta^2)=(0,8\pi)$.
We can then conclude that also in 2+1-QED at finite temperature there
is a BKT phase transition, with a critical line in the $(m/T,e^2/T)$
plane starting at $(m/T,e^2/T)=(\infty,8\pi)$. The critical
temperature for this transition (up to 1-loop order) can be computed from
Eq.(\ref{effb}) and Eq.(\ref{sing}) as
\begin{equation}
T_{\rm
crit.}=\frac{e^2}{8\pi}\left(1-\frac{e^2}{12\pi m}+\ldots\right)\ \ .
\end{equation}
This is the critical value of the coupling constant originally found
by Coleman in his discussion of bosonization of the massive Thirring
model ~\cite{co}.

Note that the vacuum expectation value of $A_0$ in the deconfined
phase, where the $Z$ symmetry is spontaneously broken, is
\begin{equation}
<A_0>=\frac{2\pi n T}{e}\ \ .
\end{equation}
In a semiclassical analysis, this expectation value contributes an
imaginary chemical potential for the electron action.  However, this
chemical potential can be absorbed by shifting the Matsubara
frequency.  Thus, the semiclassical thermodynamics do not suffer from
the difficulties of the meta-stable $Z_N$ phases of QCD
{}~\cite{cds,bksw}.
\\
\vspace{1 cm}

\noindent{\large\bf Aknowledgments}

G.G. and P.S. wish to thank the Physics Department of the University
of British Columbia for the hospitality and the Istituto Nazionale di
Fisica Nucleare, Sezione di Perugia, for financial support. G.S.
thanks the University of Perugia for their kind hospitality, a NATO
Scientific Exchange Grant, the Istituto Nazionale di Fisica Nucleare
and the Natural Sciences and Engineering Research Council of Canada
for financial support.

\end{document}